\documentstyle[12pt]{article}
\begin{document}
\centerline{\bf Classical Phase Space Structure Induced by Spontaneous
Symmetry Breaking  }
\vspace{1cm}

\centerline{  Marius Grigorescu}

\vspace{2cm}

\noindent
Abstract: {\small 
The collective dynamics of a many-body system is described as a
special case of low-energy quantum dynamics, occurring when the 
ground state breaks a continuous symmetry of the Hamiltonian.
This approach is applied to the spontaneous breaking  of the rotational
symmetry of a nuclear Hamiltonian. It is shown that the excitation
operator of the isovector low-lying angular oscillations in deformed
nuclei is a linear combination between angular momentum operators, which 
generate static rotations, and "angle" operators, which generate the
transition to a rotating frame.  }     
\\[1cm]
 \indent
The spontaneous symmetry breaking and the structure of the physical vacuum
are known as outstanding problems of the present field theory \cite{tdl}, 
but in a  similar form these problems appear also in nuclear physics. The
close analogy \cite{ri} between the algebraic structure of the Poincar\'e
transformation group of the space-time coordinates, and of the group
CM(3), of the nuclear collective coordinates, suggests that in both cases
we face the same basic phenomenon consisting in the occurrence of
classical structures at quantum level, but at a different energy scale. \\
\indent
A localized, or deformed nuclear mean-field indicates the breaking of the
translational, or rotational symmetry in the many-body ground state. The
corresponding Goldstone bosons appear in this case as the "spurious modes"
of the random phase approximation. These modes are related to large 
amplitude collective motions of the system, and their treatment require 
to define, at least locally, canonical coordinates and momenta. \\ \indent
Here this problem is solved by considering  the collective dynamics as a
special case of low-energy quantum dynamics occurring when the quantum
ground state $\vert g>$ is not invariant to the action of a continuous
symmetry group ${\cal G}$ of the Hamiltonian. In the many-body case this
situation appears often for the ground-state obtained by
Hartree-Fock (HF), or Hartree-Fock-Bogolyubov (HFB) calculations. These
calculations lead to a ground state $\vert g^M>$, which is a critical
point of minimum energy for the classical system obtained by constraining
the  quantum dynamics from the infinite dimensional many-body Hilbert
space ${\cal H}$, to some finite dimensional manifold $M$ of trial
functions. This is assumed to be a ${\cal G}$ - invariant symplectic
manifold $( M, \omega^M)$ with $\omega^M$ the symplectic form defined
by restricting to $M$ the natural symplectic form on ${\cal
H}$,  $\omega^{\cal H}_\psi = 2 \Im < X
\vert Y>$, $X,Y \in T_\psi {\cal H}$ \cite{am}. \\ \indent
If $\vert g^M>$ is a symmetry breaking ground state, then a whole 
critical submanifold $Q \subset M$, $Q= {\cal G} \cdot \vert g^M> $ may be
generated by the action of ${\cal G}$. Suppose that $\omega^M$ vanishes on
the tangent space $TQ$ of $Q$, if $Q$ is an isotropic submanifold of $M$.
When the algebra $g$ of ${\cal G}$ is semi-simple, this implies that the
representation operators for $g$ have vanishing expectation values in the
state $\vert g^M>$. In particular, for a deformed ground state the 
average of the angular momentum operators should be zero, and $Q$ 
appears as the coordinate space for the rotational collective degrees of
freedom.
\\
\indent
If $Q$ is isotropic, then at any $q \in Q$ the tangent space $T_q Q$ 
has a coisotropic $\omega^M$ - orthogonal complement $F_q$, and the
quotient $E_q = F_q / T_q Q$ is a symplectic vector space \cite{am}. Let
$P_q$ be the complement of $F_q$, such that $T_qM=P_q+F_q$. Then $P_q$ is
isotropic, with the same dimension as $T_qQ$, and such that 
$\omega^M$ restricted to $P_q \times T_qQ$ is non-degenerate. Thus,
locally, at every point $q \in Q$ one has a classical phase-space 
structure, with $P_q$ representing the space of the momenta canonically
conjugate to the collective coordinates. The remaining "intrinsic"
variables are represented within $E_q$. \\ \indent   
A simple and instructive application of this geometrical construction
concerns the non-trivial case of the canonical momentum associated to a
single angle coordinate, $\phi$. Let ${\cal G}_1$ be the group of
rotations around the $X$ axis generated by the total orbital angular momentum
operator $L_1$, $\vert g^M>$ a deformed ground state, and $J:M \rightarrow
R$ the momentum mapping $J( \vert Z> ) = < Z \vert L_1 \vert Z>$. Then for
any  regular value $I$ of $J$, ${\cal F}_I = J^{-1} (I)$ is invariant to
the action of ${\cal G}_1$ and $ Q = {\cal G}_1 \cdot \vert g^M > \subset
{\cal F}_0$. Moreover, $F_q = T_q {\cal F}_0$, and a 1-dimensional
complement $P_q$ for ${\cal F}_q$ in $M$ is provided by the tangent to
any curve transversal to ${\cal F}_0$ at $q$. The ambiguity in the choice
of the transversal may be solved by using dynamical arguments. An unique
"yrast" transversal is defined by joining continuously the minimum energy
points from each ${\cal F}_I$ near ${\cal F}_0$. Such constrained minima 
of the energy function $< Z \vert H \vert Z>$, $\vert Z> \in M$, 
can be obtained by a standard variational calculation for the 
"cranking Hamiltonian" $H'= H - \omega L_1$ \cite{rs}, where $\omega$ is 
the Lagrange multiplier.  Thus, the constrained minimum is a solution
$\vert Z>_\omega$ of the variational equation 
\begin{equation}
\delta <Z \vert H - \omega L_1 \vert Z> =0~~,~~ \vert Z> \in M
\end{equation}
with $\omega$ fixed at the value $\omega_I$ given by the implicit
equation $J( \vert Z>_{ \omega_I} ) = I$. The result is a symplectic
manifold 
\begin{equation}
{\cal S} = \{ \vert Z >_{( \phi,I)} = e^{ - i \phi L_1} \vert Z>_{
\omega_I} \}
\end{equation}
parameterized by the canonical variables $\phi$ and $I$. \\ \indent
If $H$ is a nuclear Hamiltonian consisting of a single-particle spherical
oscillator term and a quadrupole-quadrupole (QQ) interaction, and $M$ is
the trial manifold of the HF states, then $\vert Z>_\omega$ is a Slater 
determinant constructed with the cranked anisotropic oscillator
eigenstates. These eigenstates are connected with the spherical harmonic
oscillator eigenfunctions by the unitary operator
\begin{equation}
{\cal U}_\omega  =  e^{ -i  \lambda c_1} 
e^{-i   \sum_{k=1}^3 \theta_k s_k }
\end{equation}
with 
\begin{equation}
c_1= b^\dagger_2 b_3 + b^\dagger_3 b_2~~,~~s_k= i(b^\dagger_k b^\dagger_k
-  b_k b_k)/2 ~~,
\end{equation}
$ b^{\dagger}_k = \sqrt{ m \omega_0/ {2} \hbar} (x_k - i p_k /{m \omega_0})$, 
$\omega_0^2 = ( \omega^2_2 + \omega^2_3)/2$,
$\tan 2 \lambda = 2 \omega / \omega_0 \eta$, 
$\sinh  2 \theta_k = \omega_0 (1-\omega_k^{2} / \omega_0^2)/2 \Omega_k$.
Here $\omega_k$ are the anisotropic oscillator frequencies,
$ \eta=(\omega^{2}_2 - \omega^{2}_3)/2 \omega^2_0$,  $\Omega_1= \omega_1$,
$ \Omega_{2,3}^{2} =  (\omega_0 + \epsilon_{2,3})^{2}- (\omega_0 \eta /2)^{2}$,
$\epsilon_2= - \epsilon_3 = \omega_0 \eta /2 \cos 2 \lambda$.
\\ \indent
The operators $s_{1,2,3}$ generate the transition from a "spherical" to a
"deformed" basis, while $c_1$ appears as an "angle" operator because it
generates a shift in the expectation value of the angular momentum. 
\\ \indent
A direct application of this angle operator is provided by the treatment
of the nuclear low-lying isovector "scissors" vibrations in deformed
nuclei, considered in the study  of nuclear magnetism. Constructing
manifolds ${\cal S}_p$ and ${\cal S}_n$ separately for protons and
neutrons, and using their direct product as trial function in a 
time-dependent variational calculation for a microscopic Hamiltonian 
including both isovector and isoscalar QQ interaction terms \cite{mg}, it
can be shown that the excitation operator for the scissors modes  is
\begin{equation}
B^\dagger = \frac{1}{2} [ a^p L^p_1 - a^n L^n_1 -
\frac{i \Omega}{ \omega_2 - \omega_3} ( a^p C^p_1 - a^n C^n_1) ]
\end{equation}
where $\Omega$ is the scissors mode frequency, $C^{p,n}_1=
\sum_{i=1}^{Z,N} c^i_1$, and $a^{p,n}$ are the quantized angular
amplitudes \cite{mg}.  

\end{document}